\documentclass[twocolumn]{aastex631}

\usepackage{graphicx}
\usepackage[flushleft]{threeparttable}
\usepackage{amsmath}	% Advanced maths commands
\usepackage{amssymb}	% Extra maths symbols
\usepackage{xcolor}

\newcommand{\s}{\,{\rm s}}      
\newcommand{\ps}{\,{\rm s}^{-1}}

\newcommand{\cm}{\,{\rm cm}}    
\newcommand{\km}{\,{\rm km}}

\newcommand{\K}{\,{\rm K}}

\newcommand{\snr}{SN~1006}

\newcommand{\RAdot}[4]{{#1}^{{\rm h}}{#2}^{{\rm m}}{#3}\fs{#4}}
\newcommand{\decldot}[4]{{#1}^{\circ}{#2}'{#3}\farcs{#4}}

\newcommand{\decl}[3]{{#1}^{\circ}{#2}'{#3}''}

\begin{document}

\title{Magnetic structures and turbulence in SN 1006 revealed with imaging X-ray polarimetry}

\correspondingauthor{Ping Zhou}
\email{pingzhou@nju.edu.cn}
\author[0000-0002-5683-822X]{Ping Zhou}
\affiliation{School of Astronomy and Space Science, Nanjing University, Nanjing 210023, PR China}
\author[0000-0001-6511-4330]{Dmitry Prokhorov}
\affiliation{Anton Pannekoek Institute for Astronomy \& GRAPPA, University of Amsterdam, Science Park 904, 1098 XH Amsterdam, The Netherlands}
\author[0000-0003-1074-8605]{Riccardo Ferrazzoli}
\affiliation{INAF Istituto di Astrofisica e Planetologia Spaziali, Via del Fosso del Cavaliere 100, 00133 Roma, Italy}
\author[0000-0001-9108-573X]{Yi-Jung Yang}
\affiliation{Department of Physics, The University of Hong Kong, Pokfulam, Hong Kong}
\affiliation{Laboratory for Space Research, The University of Hong Kong, Hong Kong}
\author[0000-0002-6986-6756]{Patrick Slane}
\affiliation{Center for Astrophysics | Harvard \& Smithsonian, 60 Garden St, Cambridge, MA 02138, USA}
\author[0000-0002-4708-4219]{Jacco Vink}
\affiliation{Anton Pannekoek Institute for Astronomy \& GRAPPA, University of Amsterdam, Science Park 904, 1098 XH Amsterdam, The Netherlands}
\author[0000-0002-8665-0105]{Stefano Silvestri}
\affiliation{Istituto Nazionale di Fisica Nucleare, Sezione di Pisa, Largo B. Pontecorvo 3, I-56127 Pisa, Italy}
\author[0000-0002-8848-1392]{Niccolò Bucciantini}
\affiliation{INAF Osservatorio Astrofisico di Arcetri, Largo Enrico Fermi 5, 50125 Firenze, Italy}
\affiliation{Dipartimento di Fisica e Astronomia, Università degli Studi di Firenze, Via Sansone 1, 50019 Sesto Fiorentino (FI), Italy}
\affiliation{Istituto Nazionale di Fisica Nucleare, Sezione di Firenze, Via Sansone 1, 50019 Sesto Fiorentino (FI), Italy}
\author{Estela Reynoso}
\affiliation{Instituto de Astronom\'{i}a y F\'{i}sica del Espacio (IAFE), Av. Int. Guiraldes 2620, Pabell\'{o}n IAFE, Ciudad Universitaria, 1428 Ciudad Aut\'{o}noma de Buenos Aires, Argentina}
\author{David Moffett}
\affiliation{Department of Physics, Furman University, Greenville, SC 29613, USA}
% ...


\begin{abstract}

 Young supernova remnants (SNRs) strongly modify surrounding magnetic fields, which in turn play an essential role in accelerating cosmic rays (CRs). X-ray polarization measurements probe magnetic field morphology and turbulence at the immediate acceleration site. We report the X-ray polarization distribution in the northeastern shell of SN1006 from a 1~Ms observation with the Imaging X-ray Polarimetry Explorer (IXPE). We found an average polarization degree of $22.4\pm 3.5\%$ and an average polarization angle of $-45.4\pm 4.5^\circ$ (measured on the plane of the sky from north to east).
 The X-ray polarization angle distribution reveals that the magnetic fields immediately behind the shock in the northeastern shell of SN~1006 are nearly parallel to the shock normal or radially distributed, similar to that in the radio observations, and consistent with the quasi-parallel CR acceleration scenario.
 The X-ray emission is marginally more polarized than that in the radio band.
 The X-ray polarization degree of SN~1006 is much larger than that in Cas~A and Tycho,  together with the relatively tenuous and smooth ambient medium of the remnant, favoring that CR-induced instabilities set the turbulence in SN~1006 and CR acceleration is environment-dependent.

\end{abstract}


\section{Introduction} \label{sec:intro}

Supernova remnants (SNRs) are regarded as factories of relativistic particles.
\cite{baade34} first proposed supernovae as a potential origin of cosmic-ray (CRs) in the 1930s.
The first observational evidence for their hypothesis was the detection of radio emission from SNRs.
The radio emission is found to be non-thermal, generated by the relativistic electrons gyrating in the magnetic fields \citep[e.g.,][]{vanderlaan62}. 
Nevertheless, it lacked direct evidence for SNR accelerating CRs up to energies of $100$~TeV energy until the discovery of non-thermal X-ray emission from bilateral shells of \snr\ \citep{koyama95}.
In the high-energy regime, H.E.S.S and Fermi-LAT observations show that SN~1006 has a shell-like morphology in the  TeV and GeV band \citep{acero10,xing16,condon17}, further confirming the particle acceleration in the shock front.
X-ray synchrotron emission has also been detected in a few other young SNRs that drive fast shocks with speeds of a few thousand km~s$^{-1}$.
As synchrotron emission is produced by relativistic electrons gyrating in the magnetic fields, it keeps crucial information on particle acceleration, radiative processes, and magnetic fields. 

Two key questions in the study of CRs are how magnetic fields influence CR acceleration and how SNRs and their CRs modify magnetic fields.
The diffusive shock acceleration (DSA) theory has predicted that magnetic fields can be strongly amplified through either the streaming instability \citep{bell04} or turbulent dynamo progress via density fluctuations \citep[e.g.,][]{giacalone07}.
This prediction is supported by the presence of thin X-ray synchrotron filaments and rapid variability of nonthermal X-ray spots in young SNRs, where the magnetic fields are found to be orders of magnitude stronger than the typical value in the Galaxy \citep{vink03a, parizot06, uchiyama07}. 
The X-ray observations before the launch of IXPE had not directly revealed how the amplified magnetic fields are orientated and ordered. 
Magnetic field amplification and turbulence are suggested to occur close to the shock fronts, which are best traced by the X-ray synchrotron emission.
X-ray polarization measurements thus directly probe the magnetic field direction and turbulence, helping in understanding the mechanisms of magnetic field amplification.
The X-ray polarization angle (PA; measured on the plane of the sky from north to east; International Astronomical Union standard) orients perpendicularly to the magnetic fields, while the polarization degree (PD) reflects the turbulence level, with more ordered magnetic fields probed with a high PD.

IXPE \citep{weisskopf21} opened the imaging X-ray polarization window to probe magnetic field distribution and turbulence in high-energy sources, including SNRs.
The very recent X-ray polarization measurements of the young SNRs Cas A \citep{vink22} and Tycho \citep{ferrazzoli23} showed radially distributed magnetic fields. 
Cas~A has a relatively low polarization degree PD$\sim 5\%$ at the outer rim, while Tycho is more polarized with PD$\sim 12\%$. The difference in PD suggests a different turbulence level or scale between these two SNRs, but both are smaller than the theoretical maximum value of PD$_{\max}=\Gamma/(\Gamma+2/3)\sim 80\%$, with $\Gamma$ the photon index.

\snr\ (G327.6$+14.6$) was the third SNR observed with IXPE. It is a Galactic SNR with a diameter of $\sim 0.5^\circ$ at a distance of $2.18\pm 0.08$~kpc \citep{winkler03}. 
The IXPE observation covered the SNR's northeastern shell, which is composed of two rims (see circles in Figure~\ref{fig:fov}).
This remnant has some unique properties for studying magnetic turbulence and CR acceleration.
Firstly, \snr\ is among the best sources for probing spatially resolved magnetic properties since its nonthermal rims with typical widths of $\sim 20$--30$''$ \citep{bamba03,ressler14} can be roughly resolved by IXPE with an angular resolution of $\sim 30''$ \citep{weisskopf21}.
The bright nonthermal rims are produced by TeV electrons in magnetic fields amplified
to tens of the Galactic level \citep{parizot06}, and are
not contaminated by thermal emission in 2--8 keV.
Secondly, \snr\ evolves in a tenuous ambient medium high above the Galactic plane ($\sim 560$~pc), in contrast to Cas~A and Tycho, which are both located in the Galactic plane. The density fluctuation of the northeastern and eastern shells appears small, explaining the faint and fairly uniform surface brightness of the H$_\alpha$ emission \citep{winkler03,giuffrida22}. 
\snr\ is thus a test bed to study magnetic turbulence driven predominately by CRs, because the turbulence from small-scale density fluctuations is not considered an essential factor in this remnant.
Thirdly,
the bilateral morphology \snr\ in both X-ray \citep{koyama95} and radio \citep{reynolds86} bands has been attributed to the magnetic field orientation \citep{fulbright90,petruk09,west16}. 
\snr\ has been the focus target of observations for a
long-standing question on whether magnetic fields ``quasi-parallel'' or ``quasi-perpendicular'' to shocks are more efficient in accelerating particles \citep{bamba03, reynoso13, winkler14}.
In particular, \cite{rothenflug04} suggested that the bright X-ray synchrotron is associated with the polar caps, suggesting that there the magnetic field runs parallel to the shock normal.
A direct way to answer the question is to measure the magnetic field distribution and turbulence in \snr.

\section{Observations and data reduction}

\subsection{IXPE}

IXPE is an imaging X-ray polarimetry explorer working in the 2--8 keV band \citep{weisskopf21}. It contains three X-ray telescopes.
A detector unit (DU) is placed at the focus of each telescope, and each contains a gas-pixel detector \citep{costa01,bellazzini06} that probes the polarization of X-ray photons.
IXPE provides an angular resolution of $24''$--$30''$ (half-power diameter) and a total effective area of $\sim 76~\cm^2$ at 2.3~keV, and a field-of-view of $12\farcm{9}$ in diameter \citep{weisskopf21}, which allow for imaging polarimetry of extended sources such as SNRs.

\begin{figure}
\centering
	\includegraphics[width=0.45\textwidth]{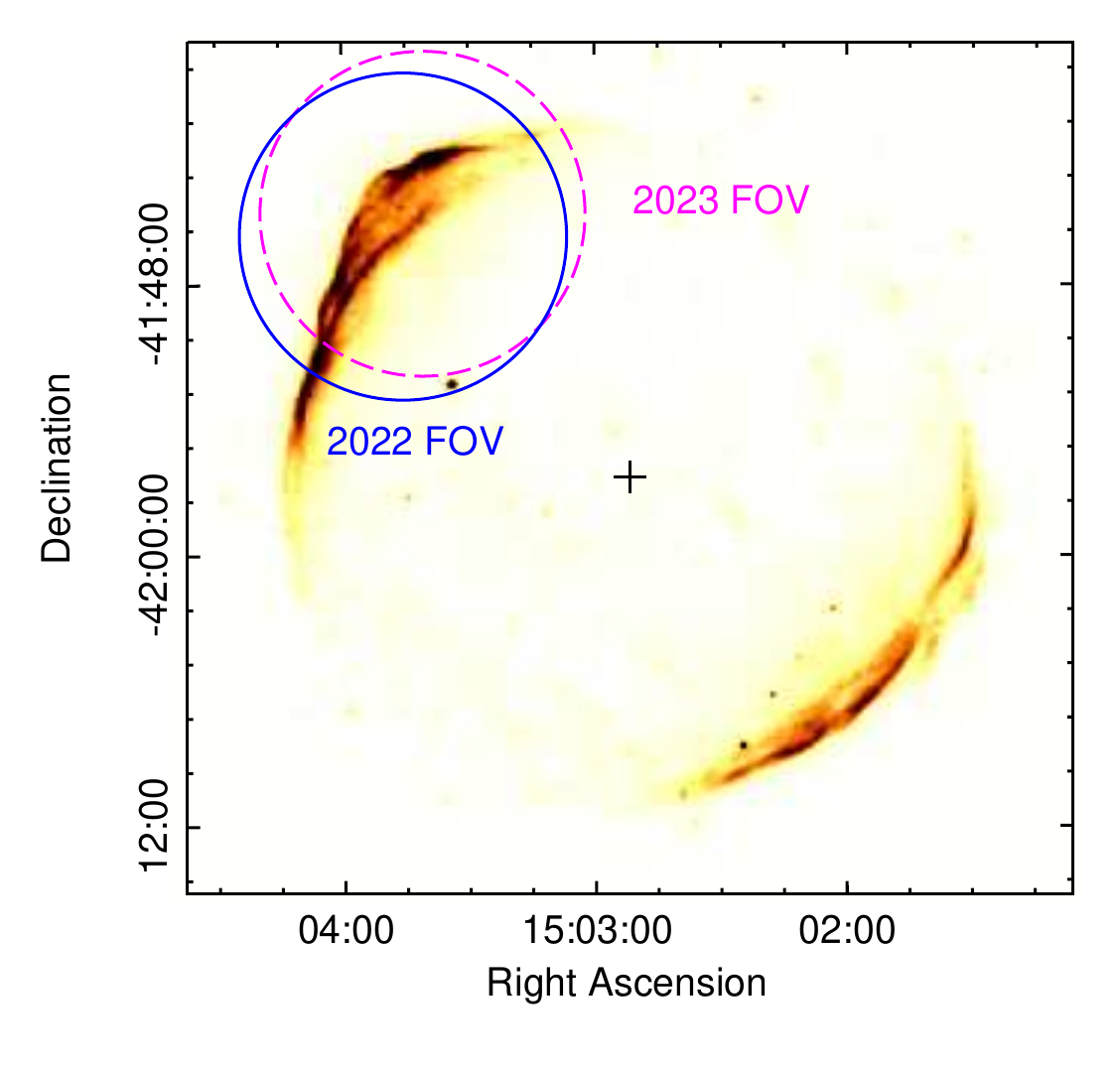}
    \caption{
    Field-of-view of the two IXPE observations overlaid on the XMM-Newton 2--8~keV image of SN~1006 \citep{li16}. The plus sign denotes the SNR center ($\RAdot{15}{02}{51}{7}$, $\decl{-41}{56}{33}$, J2000) located by \cite{reynolds86} and \cite{winkler97}.} 
    \label{fig:fov}
\end{figure}

IXPE observed the northeast shell of SN~1006 (see Figure~\ref{fig:fov}) in  2022 August 5--19 with the ID 01006801 and 2023 March 3--10 with ID 02001701, respectively. 
The livetime of the three detectors is 628--641~ks for the observation in 2022 and 339~ks in 2023.
A small fraction of observations suffered from the contamination of solar flares, which caused high photon flux in the light curves of the data.
To assess that,
we binned the 2--8~keV photons from the source region (northeastern shell)  every 120~s and obtained the light curves. To remove the time with solar flare contamination, we used a Gaussian function to fit the count rates distribution and filtered out the time intervals showing count rates higher than $3\sigma$ above the mean level. Consequently, 3--5~ks were removed in 2022 and 2-3~ks in 2023 for DU1--3.

The software {\it ixpeobssim} (vers.\ 30.0) \citep{baldnini22} was applied to reduce and analyze the IXPE data.
We used version 12 of the {\it ixpeobssim} instrument response functions of the HEASARC CALDB.
The algorithm PMAPCUBE in {\it xpbin} under the {\it ixpeobssim} software is applied to produce images of Stokes I, Q, and U, polarization degree, and polarization angle.  
A few regions of interest are selected for in-depth analysis. Further polarization analysis for binned (e.g., within given regions) data products used the PCUBE algorithm in {\it xpbin}.
We extracted spectra of Stokes I, Q, and U with PHA1, PHAQ1, and PHAU1 algorithms in {\it xpbin}, respectively.
In the spectropolarimetric analysis, we used weighted analysis and the weight response files \citep{dimarco22}.
The Stokes I spectra were grouped to reach at least 50 counts per bin.
A constant 0.4~keV energy binning is applied to the Stokes Q and U spectra.
The spectra were analyzed using $Xspec$ package (vers.\ 12.12.1) in HEASoft software.

In the polarimetric analysis, we did not correct the vignetting effect. 
The vignetting does not influence the instrumental background. The emission crossing the mirror module assemblies is affected, but the Stokes parameters I, Q, and U are subject to the vignetting in the same way,
and thus the source PD and PA are not much influenced:
PD$=\sqrt{Q^2+U^2}/I$, $Q/I=\cos(2{\rm PA})$, and $U/I=\sin(2{\rm PA})$.
Nevertheless, for large regions covering on-axis and off-axis positions,  vignetting puts more weight on the on-axis position, where the effective area is larger.

\subsection{Other data}

We retrieved the XMM-Newton EPIC data of SN~1006 taken in 2009 (obs.\ ID: 0555630301; PI: A.\ Decourchelle), with a total exposure of 125~ks.
After filtering out the intervals with heavy proton flares, the net exposure is 87/89/60 ks for the MOS1/MOS2/pn camera. The data reduction was conducted using the Science Analysis System software  \citep[SAS; vers.\ 19.1,][]{sas14}, 
and spectra were analyzed using {\it Xspec}.

For comparison purposes, we also used  the 1.4~GHz radio polarization data of SN~1006 \citep{reynoso13} taken with the Australia Telescope Compact Array (ATCA). We used the Stokes I, Q, and U radio images to calculate the degree and angle of polarization (PD$_r$ and PA$_r$) in a few regions of interest, which will be elaborated in Section~\ref{sec:radio}. 
The polarized radio emission suffered from Faraday rotation in the ionized foreground medium, which causes a rotation of the polarization angle with the square of the wavelength $\lambda^2$: 
PA$_r^{\rm obs}$=PA$_r+$RM~$(\lambda/1~{\rm m})^2$. 
Following \cite{reynoso13}, a uniform rotation measure RM=$12~{\rm rad~m}^{-2}$ is applied in this work to obtain the original polarization angle PA$_r$, although there could be local variations of RM in the SNR.

\subsection{Position correction of the IXPE data}

There was an offset of the IXPE pointing in the two epochs of observations for SN~1006. A misalignment was also found in the Cas~A observations with the order of $2\farcm{5}$ \citep{vink22} and also in one of Tycho SNR observations \citep{ferrazzoli23}. The focal pointing ($x$, $y$) has an offset due to the boom bending and the difficulty of the bending correction for extended sources \citep{weisskopf21}. We did not consider the rotation angle correction, as it is small compared to the angular resolution of IXPE. The roll angle of IXPE is determined to an accuracy of $0.7^\circ$. Across the field of view with a diameter of $12\farcm{9}$, the maximum shift in the position of an X-ray source due to roll angle error is $4\farcs{7}$.

To correct for the positions of the two-epoch IXPE observations, we simulated a reference image using the XMM-Newton X-ray image of SN~1006 taken in 2009 (obs. ID: 0555630301; PI: A. Decourchelle). We are aware that the SNR shell has a proper motion of $\sim 0\farcs{48}$~yr$^{-1}$ \citep{katsuda09}, which can cause a northeastern motion of $\sim 6''$ for the shell in a baseline of 13~yr. Nevertheless, given the angular resolution of $30''$ for IXPE, such a small offset does not significantly influence our polarization studies.
From the XMM-Newton 2--8~keV image and spectrum, we simulated a 10~Ms IXPE observation with $ixpeobssim$ and obtained the reference image with a pixel size of $2\farcs{46}$. 
Then we compared the reference image, $img1$, and the observed image, $img2$, (also binned with a pixel size of $2\farcs{46}$) with an offset $(dx,dy)$ added.
The best-fit offset $(dx,dy)$ is obtained when the deviation between the two images $\sum_{i=1}^n[img2(i)-img1(i)]^2$ reaches the minimum, where $i$ and $n$ are the $i$th pixel and pixel number, respectively.
With this algorithm, we estimate the offset ($-14\farcs{8}, 0''$) for the observation in 2022 and  ($49\farcs{3}, 24\farcs{6}$) for the observation in 2023, but the offset correction could have an uncertainty. Based on a visual check, the final offset is much smaller than the angular resolution of IXPE.

\subsection{Background analysis}

The background analysis is crucial for faint or extended sources observed with IXPE \citep{xie21,ferrazzoli23,dimarco23}. 
SN~1006 is the faintest among the three SNRs observed so far by IXPE, and thus the background contribution needs to be considered. 
The background photons have two contributions:  the diffuse X-ray background near SN~1006 and the particle-induced instrumental background, which is nearly uniform across the detector. 
Since SN~1006 is located at the high Galactic latitude ($b=14\fdg{6}$), the Galactic diffuse background is small.
The 2--8 keV background level of previous IXPE observations was found in the range 8.7--$12.1 \times 10^{-4}$ counts s$^{-1}$ arcmin$^{-2}$ \citep{dimarco23}. 
We conducted an energy-dependent background rejection by selecting the events with the event-track region of interest following the strategy by \cite{dimarco23}. This removes up to 40\% instrumental background events.

The residual background is still non-negligible, so we selected a source-free region outside SN~1006 to estimate the background (see the dashed regions in Figure~\ref{fig:polarplot}).  Emission in this region includes the local (sky) background of SN~1006 and residual instrumental background.
We obtained the background level of 5.1--$5.9\times 10^{-4}$ counts s$^{-1}$  arcmin$^{-2}$ in 2--8 keV in DU1--3, while at the SNR shell, we obtained 1.9--$2.0\times 10^{-3}$  counts s$^{-1}$  arcmin$^{-2}$.
Therefore, the background contributes $\sim 30$\% of the photons from the shell region.

\begin{figure*}
\centering
	\includegraphics[width=0.35\textwidth, angle=0]{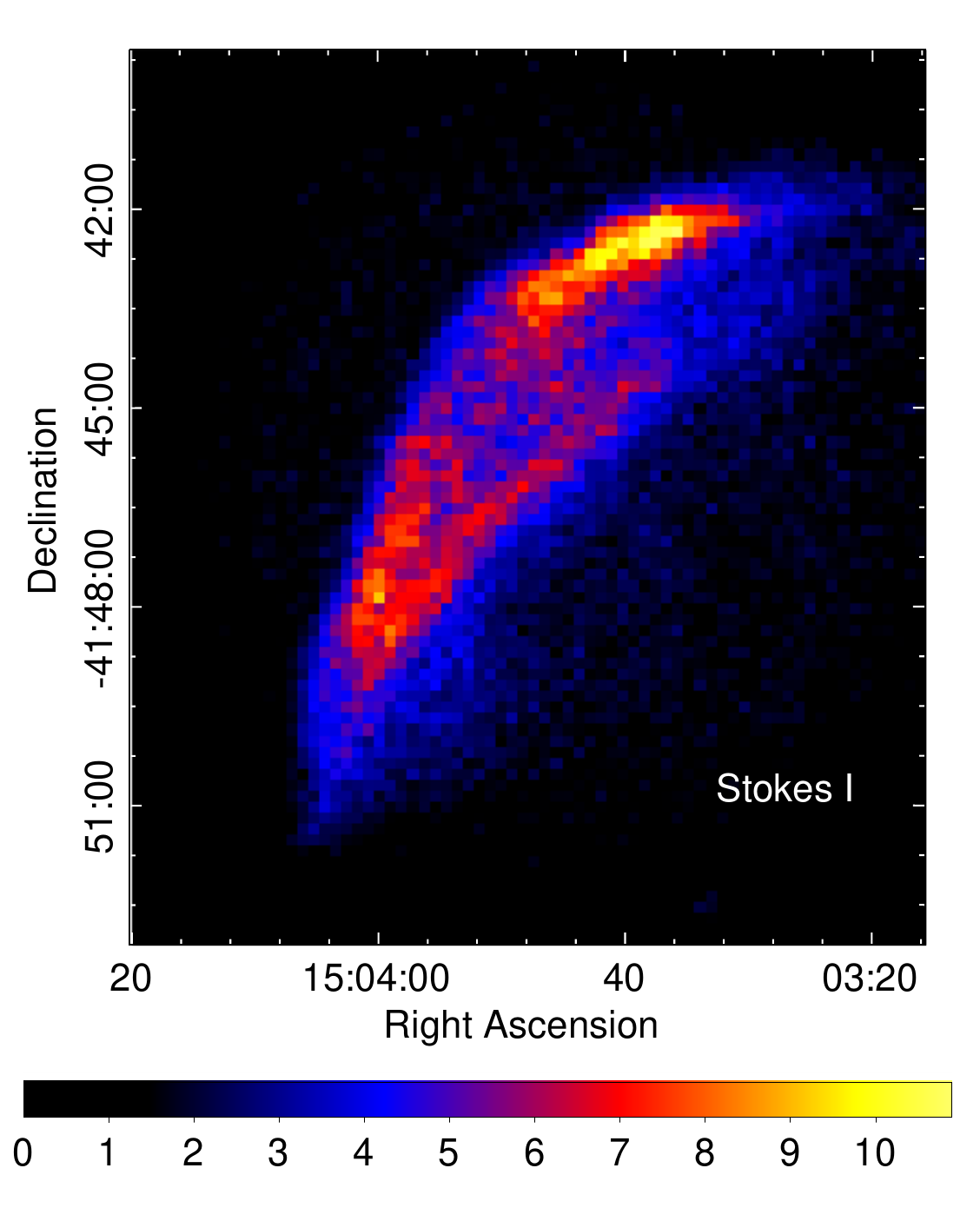}
	\includegraphics[width=0.35\textwidth, angle=0]{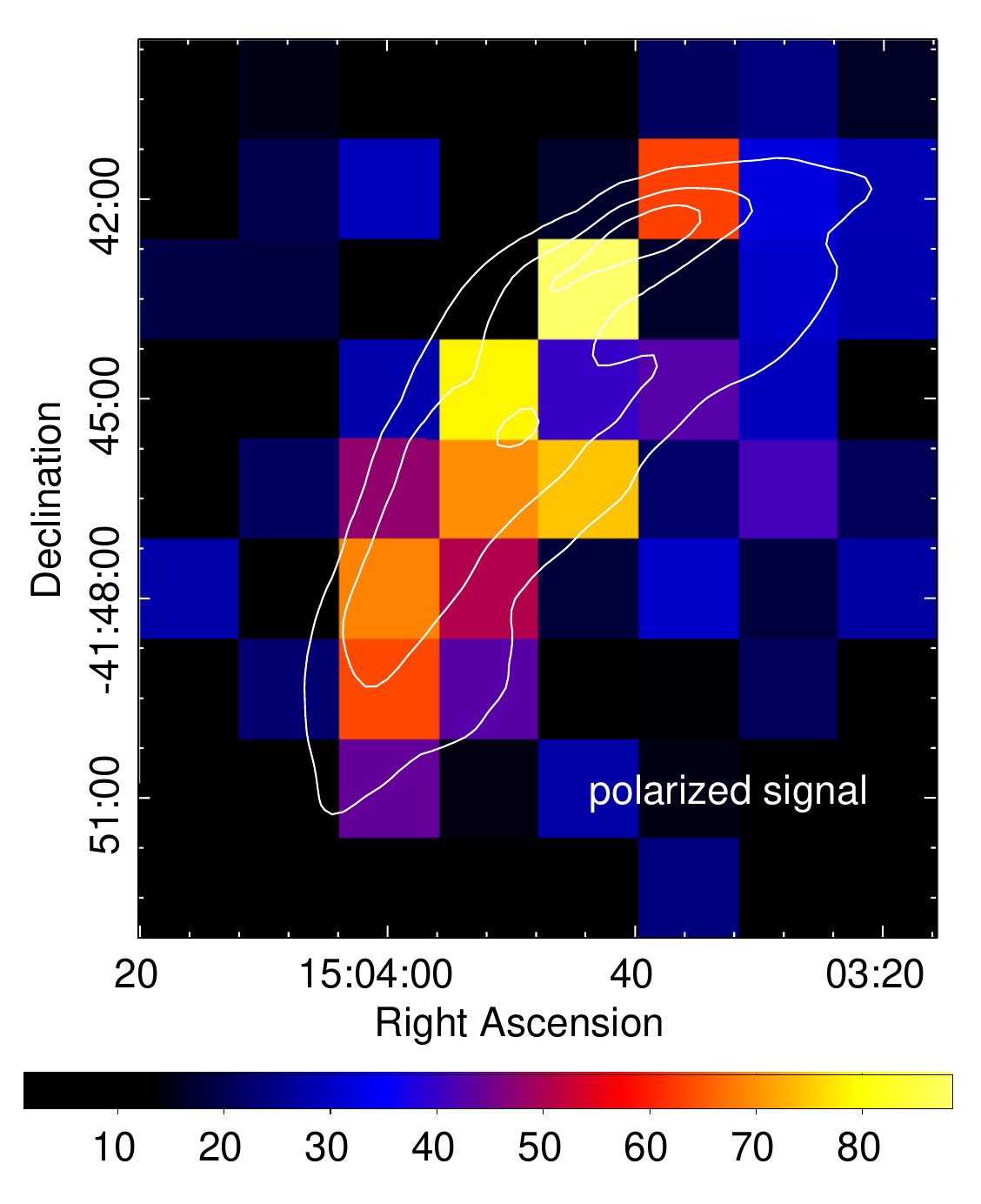}
	\includegraphics[width=0.35\textwidth, angle=0]{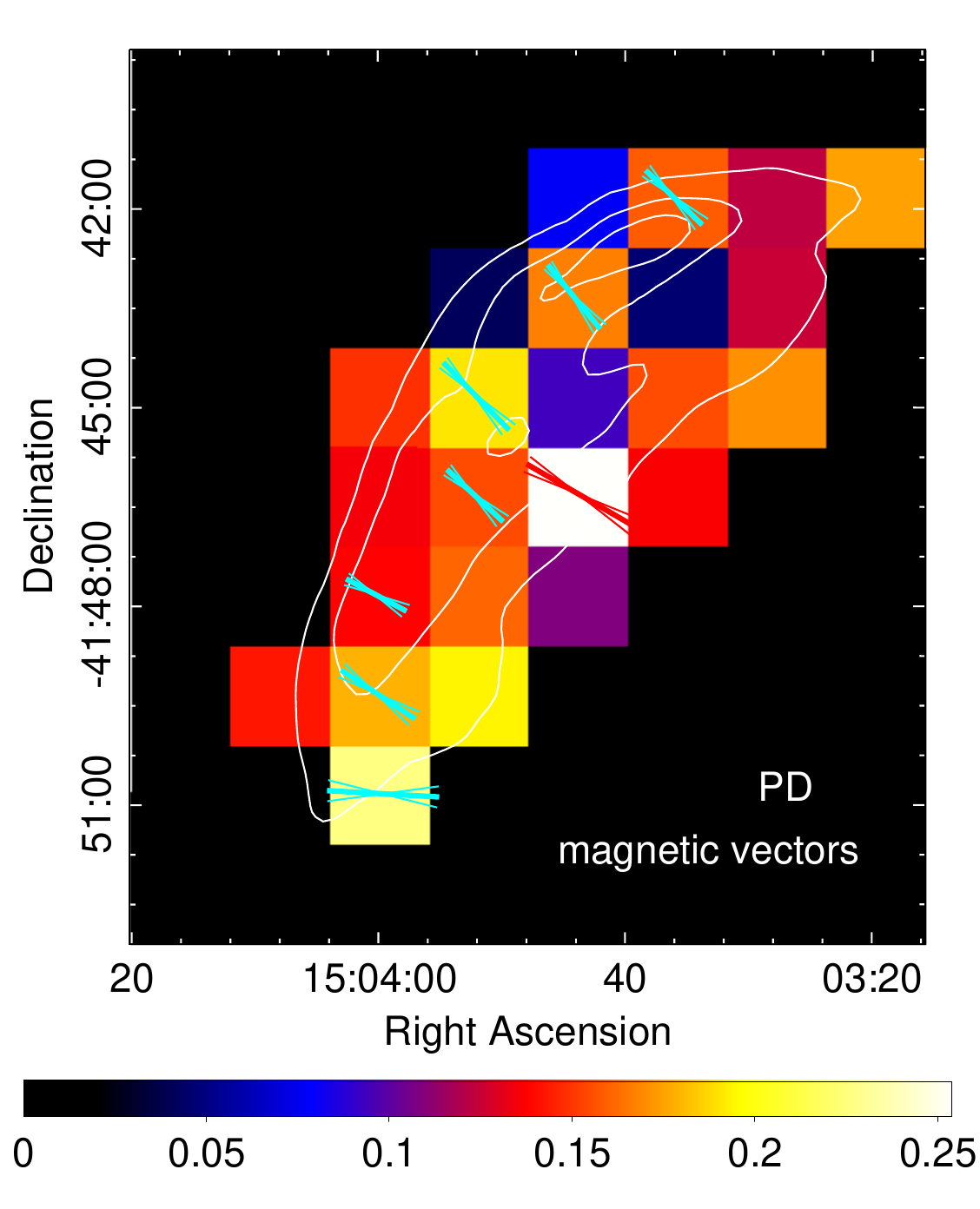}
	\includegraphics[width=0.35\textwidth, angle=0]{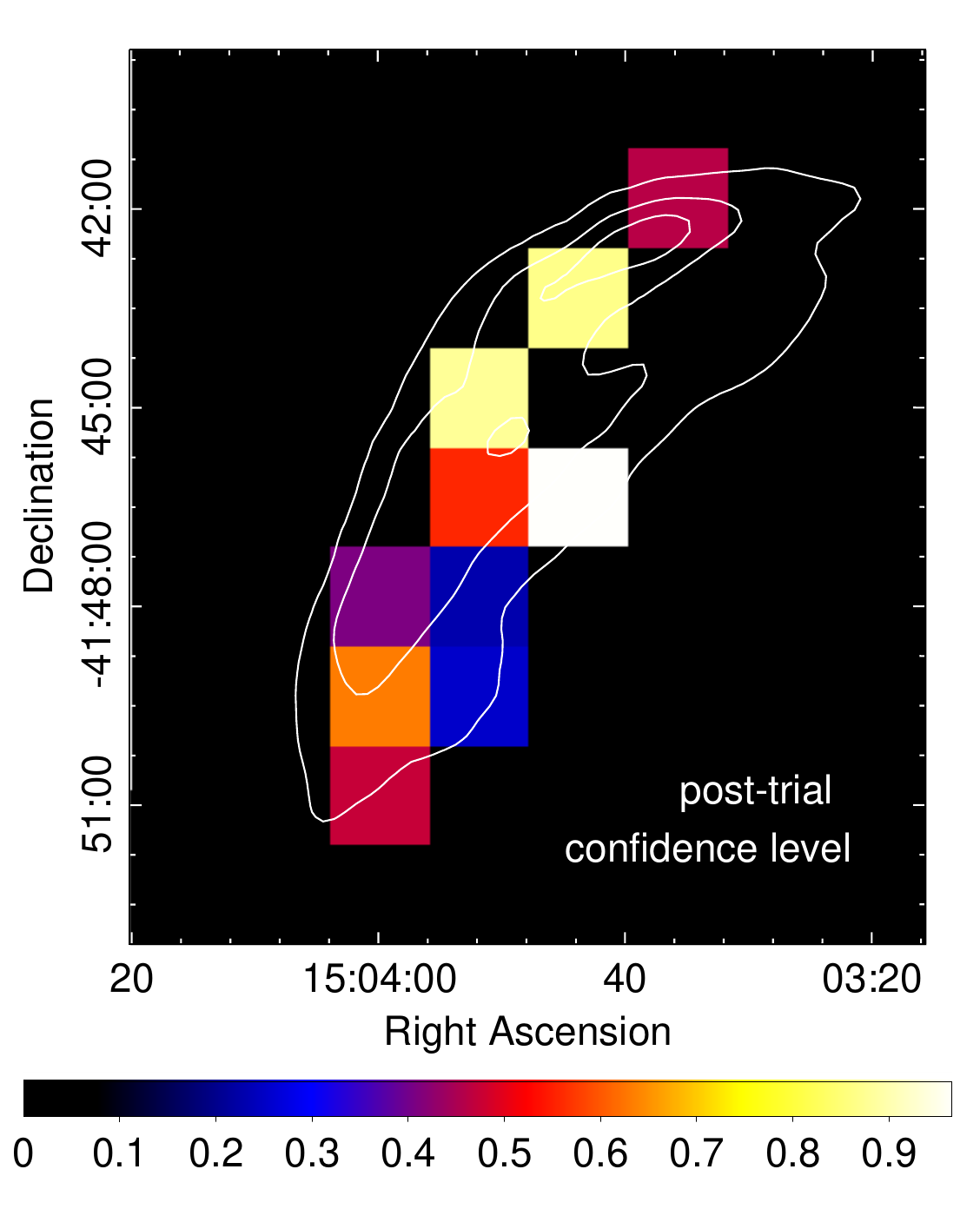}
    \caption{
    IXPE images of the northeastern shell of SN~1006 in the 2--4 keV band.
    Upper left: Stokes I image with a pixel size of $10''$. Upper right: polarized signal ($\sqrt{Q^2+U^2}$; right) image binned with a pixel $1\farcm{5}$. Lower left: polarization degree (PD) distribution for pixels with Stokes I$>158$~counts (to mask out pixels with low X-ray brightness), overlaid with magnetic vectors and their $1~\sigma$ errors. The blue and red vectors correspond to pixels with a pre-trial significance of 2--$3~\sigma$ and $>3~\sigma$, respectively. The length of the magnetic vectors scale with the polarized degree. Lower-right: the post-trial confidence level for the detection of the polarized signal. The white contours show the Stokes I levels. }
    \label{fig:i_pf}
\end{figure*}

\iffalse
\begin{figure*}
\centering
	\includegraphics[width=0.8\textwidth, angle=0]{pd_merge_err.pdf}
    \caption{
   IXPE polarization degree (PD) images of SN~1006 northeast in the 2--4 keV band, overlaid with magnetic vectors and their $1\sigma$ errors. The contours are taken from the Stoke I image (see Figure~\ref{fig:i_pf}. Only pixels with significance above $2\sigma$ and Stokes I larger than ${\rm 70(pixel~size/1~arcmin)^2}$ are shown. The blue vectors denote the pixels with $3\sigma$ significance. .}
    \label{fig:pd}
\end{figure*}
\fi

\section{Results}

\subsection{Polarization images}

IXPE  allows a spatially resolved polarization study of the double-rim structure in the northeast shell of SN 1006.
Figure~\ref{fig:i_pf} (left) shows the Stokes I count morphology in the 2--4~keV band with a pixel size of $10''$. 
We selected the 2--4~keV energy band to minimize the contamination from the strong instrumental background above 4 keV (see Appendix and Figure~\ref{fig:spec}). 
The distribution of polarized X-ray emission in the northeast of SN1006 with a bin size of $1\farcm{5}$ is also shown in Figure~\ref{fig:i_pf} (right). 
The polarized emission is made using the Stokes Q and U maps ($\sqrt{Q^2+U^2}$) and generally follows the distribution of Stokes I. 

The PD distribution in the shell is shown in the lower-left panel of Figure~\ref{fig:i_pf} (pixel size of $1\farcm{5}$). The magnetic vectors and the $1~\sigma$ uncertainty range are overlaid on pixels with a pre-trial confidence level ${\rm CL_{pre}}>95\%$ ($2~\sigma$).
In these pixels, the PDs range from 14\% to 25\%, and the uncertainty is $\approx 6\%$, except for the southern pixel with a larger error of $9\%$.
%The PA values are between $-51^\circ$  and $-3^\circ$ {\bf (the northern pixel)} and the error is 8-$11^\circ$. 
The PA values for most of these pixels are around $-45^\circ$ with an error of 8--$11^\circ$, except for the southern pixel with PA$=-3^\circ$.
The overall magnetic orientation tends to be parallel to the shock normal, consistent with the radio polarization measurement \citep{reynoso13}.

The quoted significance level per pixel is pre-trial significance, not corrected for the number of pixels.  Each independent pixel contributes to a trial, and as the number of trials (pixels) increases, the probability of finding a ``significant'' random fluctuation also increases. 
In the shell with $N=23$ independent pixels, the probability of finding no random deviation above a significance threshold (here taking ${\rm CL_{pre}}$) is  ${\rm CL_{post}=CL_{pre}}^N$.
Assuming a uniform polarization degree error across the shell, we obtained the post-trial confidence level distribution as shown in the lower-right panel of Figure~\ref{fig:i_pf}. The most significant pixel in the shell has ${\rm CL_{post}}=97\%$ and two more pixels show ${\rm CL_{post}}=87\%$--88\%.
To increase the significance, we need to enlarge the pixel size or select larger regions.
Moreover, the PD values are not corrected for the background, which introduces unpolarized photons that may reduce the PD values.

\begin{figure*}
\centering
	\includegraphics[width=\textwidth]{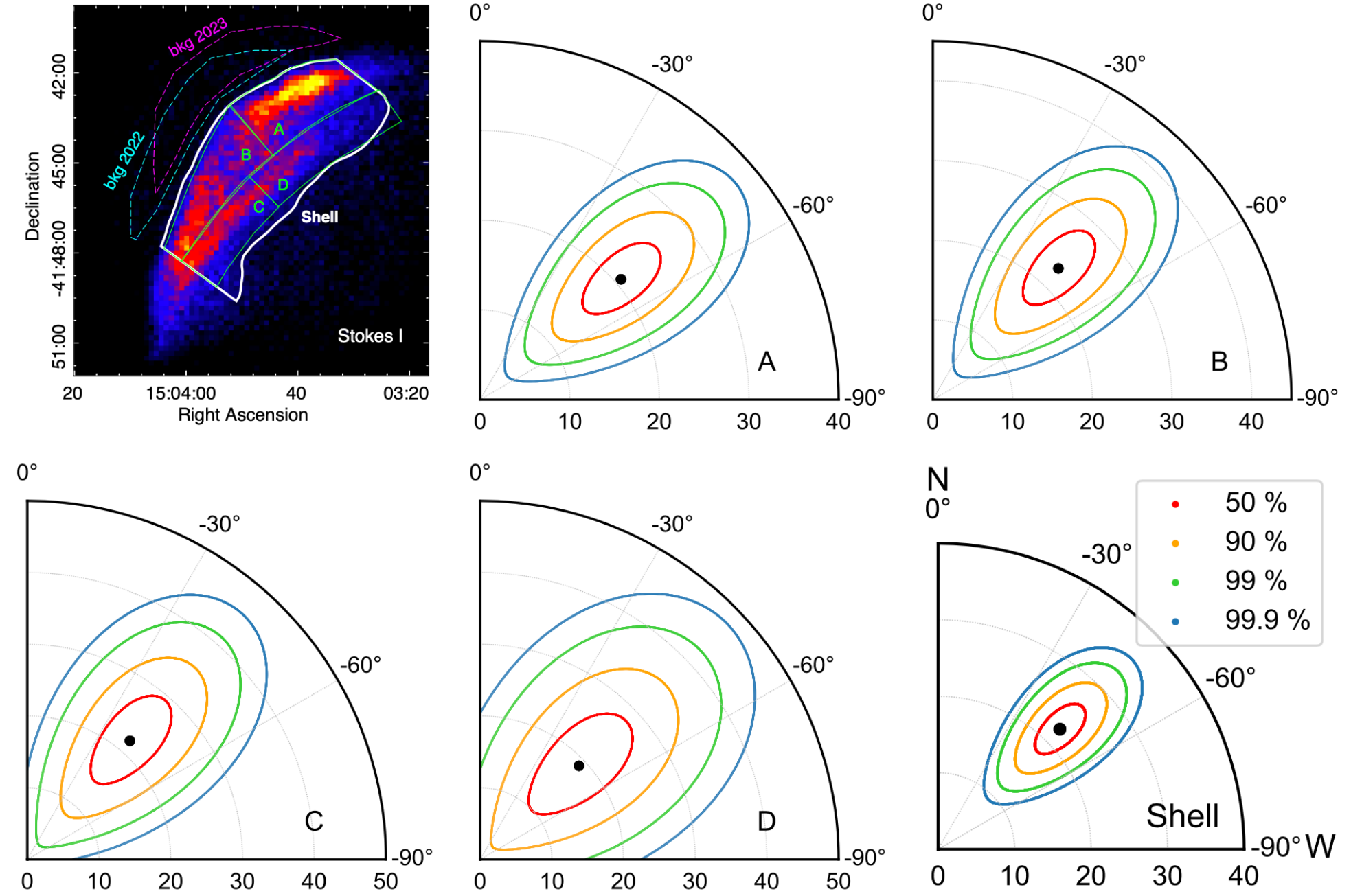}
    \caption{
    Stokes I image with regions and the polar plots of these regions in 2--4 keV band.
    In the first panel, the regions delineated with solid and dashed lines correspond to the source and background regions, respectively.
    Each polar plot shows the polarization degree (\%) and angles in the radial and position angle coordinates, respectively, with the background subtracted. The best-fit PD and PA values are denoted by black dots. The 50\%, 90\%, 99\%, and 99.9\% confidence levels (based upon $\chi^2$ on 2 degrees of freedom) are coded in red, orange, green, and blue, respectively.
}
    \label{fig:polarplot}
\end{figure*}

\iffalse

\begin{table*}
\caption{Polarization results of 5 regions by IXPE and radio observations. 
}
\centering
\begin{tabular}{l|ccccccc}
\hline
\hline
& \multicolumn{4}{c}{X-ray}  & & \multicolumn{2}{c}{radio}\\ \cline{2-5} \cline{7-8}
Region & PD  &  PA & $\sigma$ & $\Gamma$ & &  PD$_r$ & PA$_r$ \\
 & (\%) & ($^\circ$) & & & &  (\%) & ($^\circ$)   \\ \hline
Shell &  $22.4\pm 3.5$  & $-45.4\pm 4.5$ & 6.3  
& $2.74\pm 0.01$ &
& $14.5\pm 0.2$ & $-36.3\pm 0.4$\\
A &  $20.7\pm 4.5$ & $-49.5 \pm 6.2$  & 4.6 
&  $2.72\pm 0.01$&
& $12.5\pm 0.5$ & $-58.4\pm 1.0$\\
B & $22.8\pm 4.9$ & $-43.7 \pm 6.2 $ & 4.6 
& $2.68 \pm 0.01$ &
& $19.1\pm 0.6$ & $-28.2\pm 0.9$\\
C & $21.9\pm 6.4$ & $-40.0\pm 8.3$ & 3.4 
& $2.87\pm 0.01$ &
& $18.2\pm 0.4$ & $-28.0\pm 0.7$ \\
D & $<38.8$ & undetermined & 2.5 
& $3.05\pm0.02$ &
& $ 13.9\pm 0.4$ & $-39.8\pm 0.8$\\
\hline
\hline
\end{tabular}
\tablecomments{The regions are denoted in Figure~\ref{fig:polarplot}. 
The polarization errors are provided at the 1$\sigma$ confidence level.
The spectral indexes $\Gamma$ are  obtained  by fitting the XMM-Newton spectra. 
The angle of polarization in the radio band PA$_r$ is calculated using a uniform RM of $12~{\rm rad~m^{-2}}$ \citep{reynoso13}.}
\label{tab:pol}
\end{table*}
\fi

\begin{table*}
\caption{Polarization results of 5 regions by IXPE and radio observations. 
}
\centering
\begin{tabular}{l|ccc|ccc|cc}
\hline
\hline
& \multicolumn{3}{c}{IXPE (polarimetric)}  & \multicolumn{3}{c}{IXPE (spectropolarimetric)} & \multicolumn{2}{c}{radio}\\ \cline{2-4}  \cline{5-7} \cline{8-9}
Region & PD  &  PA & $\sigma$ &  PD  &  PA &  $\Gamma$ &   PD$_r$ & PA$_r$ \\
 & (\%) & ($^\circ$) & & (\%) & ($^\circ$) &  &  (\%) & ($^\circ$)   \\ \hline
Shell &  $22.4\pm 3.5$  & $-45.4\pm 4.5$ & 6.3  
& $20.2\pm 3.3$ & $-49.4\pm 4.7$ & $2.77\pm 0.04$
& $14.5\pm 0.2$ & $-36.3\pm 0.4$ \\
A &  $20.7\pm 4.5$ & $-49.5 \pm 6.2$  & 4.6 
& $21.3\pm 4.7$ & $-56.1\pm 6.2$ & $2.77\pm 0.05$
& $12.5\pm 0.5$ & $-58.4\pm 1.0$\\
B & $22.8\pm 4.9$ & $-43.7 \pm 6.2 $ & 4.6 
& $18.5\pm 4.8$ & $-44.8\pm 7.5$ & $2.65\pm 0.06$
& $19.1\pm 0.6$ & $-28.2\pm 0.9$\\
C & $21.9\pm 6.4$ & $-40.0\pm 8.3$ & 3.4 
& $27.5\pm 6.6$ & $-45.3\pm 6.5$ & $2.82\pm 0.07$
& $18.2\pm 0.4$ & $-28.0\pm 0.7$ \\
D & $<38.8$ & undetermined & 2.5 
& $<31.8$ & undetermined & $3.11\pm 0.09$ 
& $ 13.9\pm 0.4$ & $-39.8\pm 0.8$\\
\hline
\hline
\end{tabular}
\tablecomments{The regions are denoted in Figure~\ref{fig:polarplot}.  The polarization results are obtained using the 2--4~keV data and the errors
are provided at the 1$\sigma$ confidence level.
The photon indexes $\Gamma$ are  $2.74\pm 0.01$,   $2.72\pm 0.01$, $2.68 \pm 0.01$, $2.87\pm 0.01$, and $3.05\pm0.02$ for the five regions by fitting the XMM-Newton spectra in 0.8--8~keV.
The angle of polarization in the radio band PA$_r$ is calculated using a uniform RM of $12~{\rm rad~m}^{-2}$ \citep{reynoso13}.
}
\label{tab:pol}
\end{table*}

\subsection{Polarization properties of regions} \label{sec:region}

To accurately measure the polarization parameters for SN~1006, we need to consider the contribution of background photons in the source regions. 
We calculated the polarization parameters of five regions by taking the background from a nearby source-free region (see the dashed regions in Figure~\ref{fig:polarplot}).
The regions include ``Shell'' for the whole northeastern shell of SN1006, regions ``A'' and ''B'' for the outer filaments, and regions ``C'' and ``D'' for the inner filament (see Figure~\ref{fig:polarplot}).
This region division allows studying polarization differences between filaments and investigating polarization angle (PA) variation with the position angle of the structures relative to the SNR center.

The polarization degrees and angles with their 50\%, 90\%, 99\%, and 99.9\% confidence level contours of the five regions are shown as polar plots in Figure~\ref{fig:polarplot}, and the values are tabulated in Table~\ref{tab:pol}.
The polarization signal from the whole northeastern shell is detected at a significance of  $7\sigma$, while the four regions A--D are detected at the 2.5 -- 4.6$\sigma$ level.
The average polarization of the northeastern shell is $22.4\%\pm 3.5\%$, and the polarization angle is $-45\fdg{4} \pm 4\fdg{5}$.
Three smaller regions show a polarization degree of 21\% -- $23\%$, and polarization angles between $-50^\circ$ and $-40^\circ$, consistent with those of the entire shell.
The outer filaments (A and B) and the inner filament (C and D) reveal identical PD, implying a similar turbulence level of the magnetic fields. The PA in the southern regions (B and C) and northern regions (A and D) cannot be distinguished considering the error range.

Different methods exist to divide the northeastern shell and select regions. We also selected regions based on X-ray brightness distribution, as elaborated in Appendix. We found that the PD is near uniform (PD$\sim 20\%$--26\%) and does not show a correlation with the X-ray brightness.

Previous radio polarization measurements of SN~1006 found that the magnetic fields are predominantly radially distributed but also contain components nearly parallel to the Galactic plane \citep{reynoso13}.
To test whether the IXPE measurements support a radial magnetic geometry, we reconstruct events files based on the circular polarization geometry model using the $xpstokesalign$ command in $ixpeobssim$ and check whether this approach improves the detection of polarization.
The initial Stokes parameters $q_k$ and $u_k$ of each event are rotated with respect to the SNR center at R.A~(J2000)$=\RAdot{15}{02}{51}{7}$, Decl.\ (J2000)=$\decl{-41}{56}{33}$ \citep{reynolds86,winkler97}
so that the zero for the photoelectron direction is aligned with a circularly symmetric polarization model at the position of the event. 
If the PA in SN~1006 northeast has a circular symmetry rather than a uniform distribution, we would find significantly larger PD due to less depolarization.
Using the aligned events files, region ``Shell'' has PD$_{\rm align}=21.9\pm 3.5\%$ at a significance of $6.2\sigma$, similar to
that obtained from the uniform PA model (PD=$22.4\pm 3.5\%$, see Table~\ref{tab:pol}).\footnote{Similar results are obtained by taking the SNR center ($\RAdot{15}{02}{54}{9}$, $\decldot{-41}{56}{8}{0}$, J2000) measured by \cite{katsuda09}: PD$_{\rm align}=21.8\pm 3.5\%$, PA$_{\rm align}=1.5\pm 4.6^\circ$.}
This means that we cannot distinguish the radial and uniform magnetic orientation models based on current IXPE observations, which only covered the northeastern X-ray bright limb of SN~1006.
The obtained PA$_{\rm align}=0\fdg{9}\pm 4\fdg{6}$ strongly favors the radial magnetic field (PA$_{\rm align}=0^\circ$) over a tangential magnetic morphology (PA$_{\rm align}=90^\circ$).

\subsection{Spectropolarimetric analysis of regions}

A different approach to calculate the polarization of the five regions is to use the I, Q, and U spectra.
We extracted source spectra from regions ``Shell'' and  ``A--D'' and background spectra from ``bkg 2022'' for the 2022 observation and from ``bkg 2023'' for the 2023 observations (see regions in Figure~\ref{fig:polarplot}).
For each region, we jointly fit the background-subtracted I, Q, and U spectra with a model combining a foreground absorption,  a power-law spectrum, and a polarization degree and angle constant over the 2--4 keV energy band ($tbabs*powerlaw*constpol$ in XSpec).  
Since the foreground absorption can hardly be constrained with the $> 2$~keV photons, we needed to calculate $N_{\rm H}$  by fitting the XMM-Newton spectra using  the $tbabs*powerlaw$ model. 
The XMM-Newton spectral fit gives the foreground gas column density $N_{\rm H}=1.93(\pm 0.03)\times 10^{21}~\cm^{-2}$ for the shell region. Then we fixed
$N_{\rm H}$ in the IXPE spectropolarimetric analysis. 

\begin{figure*}
\centering
	\includegraphics[width=0.48\textwidth]{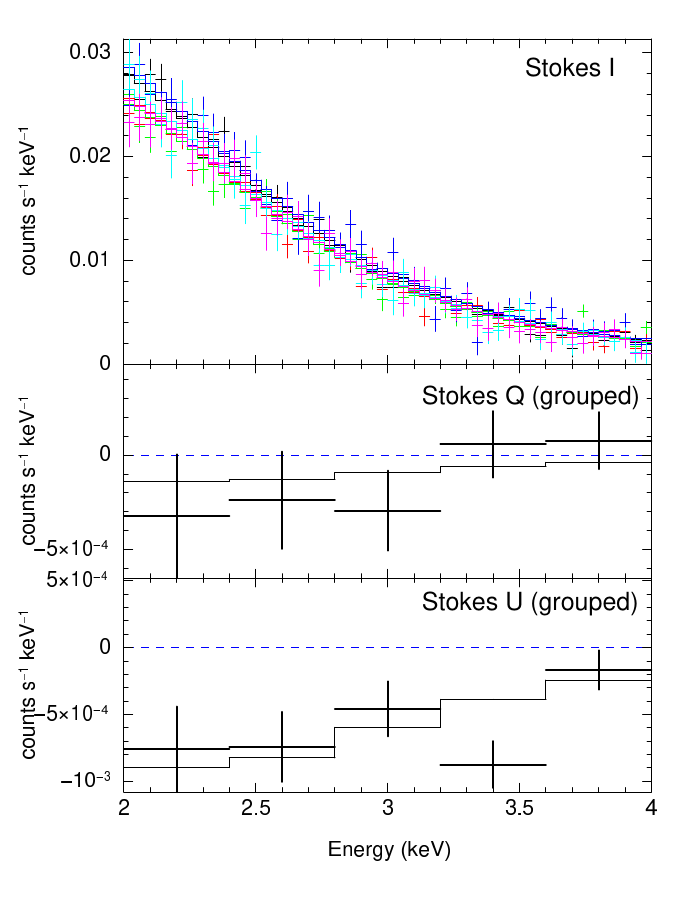}
	\includegraphics[width=0.48\textwidth]{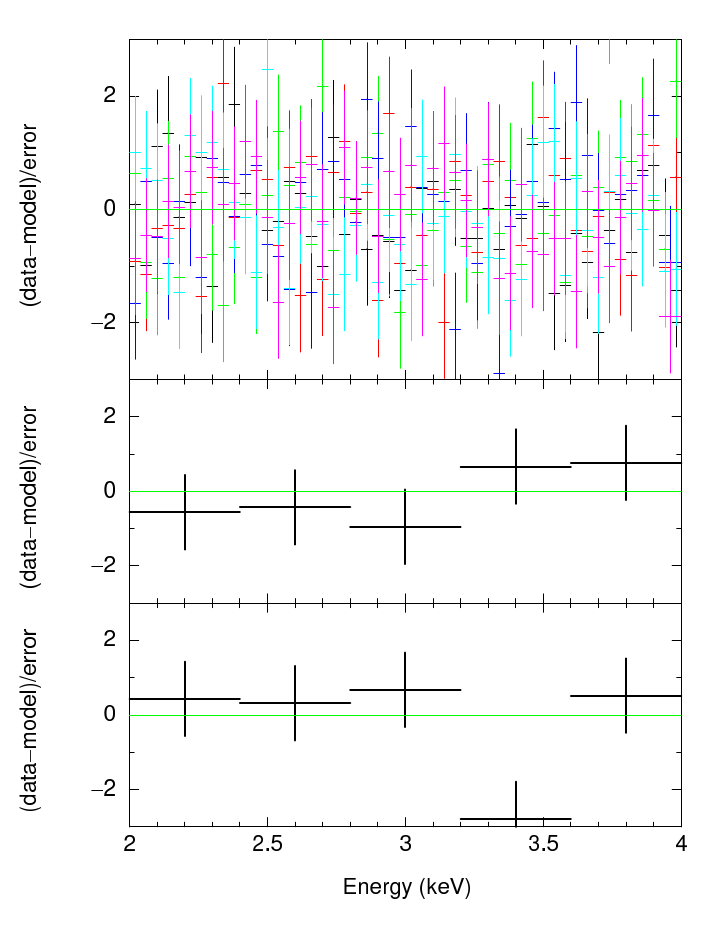}
    \caption{Left: Stokes I, Q, U spectra of the region ``Shell'' (data points) and the best-fit model (solid lines). Right:  The residuals in terms of (data$-$model)/error. The 6 Stokes I spectra from two observations are coded using different colors. The 6 Stokes Q and U spectra are grouped, respectively, for better visualization.
}
    \label{fig:spec}
\end{figure*}

Figure~\ref{fig:spec} shows the I, Q, and U spectra of region ``Shell'', the best-fit model, and the residuals.
The spectral fits produced good fit statistics with $\chi^2/dof=335.63/349=0.96$ for the ``Shell '' region, and $\chi^2/dof=0.98$--1.18 for regions A--D.
The best-fit PD, PA, and photon index $\Gamma$ in each region are shown in Table~\ref{tab:pol}. The polarization parameters obtained from the spectropolarimetric analysis agree with the polarimetric results (using PCUBE algorithm in $xpbin$). Moreover, in the ``Shell'' region, the photon index $\Gamma=2.77\pm 0.04$ of the IXPE spectra is consistent with that given by the XMM-Newton spectral fit $\Gamma=2.74\pm 0.01$. The consistency is also found for smaller-scale regions.

\section{Discussion}

\subsection{Magnetic orientation} \label{sec:orientation}

IXPE observations reveal the magnetic field aligned with the shock normal in the northeastern shell of SN~1006 (see Section~\ref{sec:region}).
The measured magnetic position angle of $44\fdg{6} \pm 4\fdg{5}$ is compatible with the position angle $24^\circ$ -- $58^\circ$ of the region ``Shell'' with respect to the SNR center (position angle $=90^\circ$ for the east of SN~1006 and $=0^\circ$ for the north). 
This is also consistent with a radial magnetic field configuration in the observed region.

The origin of radial magnetic orientation in young SNRs is still not fully understood, but a few explanations have been proposed \citep[see detailed discussions in][]{vink22}. One explanation invokes MHD instabilities introduced by density fluctuation. Young SNRs in an inhomogeneous ambient medium drive Richtmyer-Meshkov instabilities (RMI) which lead to amplifying the radial component of the magnetic fields \citep{giacalone07,inoue13}.
This scenario requires density fluctuations in the ambient medium, which could be valid for remnants in a cloudy medium.
However, SN~1006 evolves in a fairly uniform, tenuous medium 560~pc above the Galactic plane \citep{acero07,giuffrida22}.
It lacks observational evidence of a large density fluctuation for driving strong magnetic turbulence.

An alternative explanation of the radial magnetic fields in the northeast shell of SN~1006 is related to efficient particle acceleration \citep{west17}. If the CRs are more efficiently accelerated when the magnetic fields are quasi-parallel to the shock normal, the CR electron density will be amplified in the radial magnetic fields and suppressed in the tangential fields. In this case, even if the magnetic orientation is totally random, the X-ray synchrotron emission will tend to probe the enhanced CR electrons that align the radial magnetic field, resulting in an apparent radial magnetic distribution.
However, as simulated by \cite{west17}, purely random magnetic fields would predict a spherically symmetric rim, which does not explain the limb-brightened morphology of SN~1006 (see Figure~\ref{fig:fov}). 
Moreover, a problem with this explanation for the radial magnetic field orientation in the X-ray regime is that the 10--100 TeV electrons emitting X-rays have diffused away from the region from which they originated. How much of an effect that has depends on the length scales of the regions with radial magnetic-field orientations compared to the diffusion length scale \citep[see][]{vink22}.

It is likely that the pre-existing magnetic fields of SN~1006 consist of a northeast-southwest orientated component, which it kept, to some extent, after shock propagation \citep{reynoso13}. 
Our IXPE observations did not cover the whole SNR to test this hypothesis, but this was supported by the radio polarization measurement of the southeastern and northwestern portions of the shells in SN~1006 \citep{reynoso13}.
This northeast-southwest magnetic distribution is favored by  3D MHD simulations, which found that the quasi-parallel CR acceleration model best reproduces the radio morphology \citep{bocchino11} and polarization images of SN~1006 \citep{schneiter15}.

The X-ray polarization angle distribution of SN~1006 is consistent with efficient quasi-parallel CR acceleration, which enhances the CR electron density and thus the synchrotron emission in the northeastern and southwestern limbs of SN~1006. 
Moreover, turbulent magnetic fields can be generated in the quasi-parallel case and vanish in the quasi-perpendicular case \citep{caprioli14}. 
This can explain the PD$<$PD$_{\rm max}$ in the northeast shell of SN~1006.

\subsection{Magnetic turbulence} \label{sec:turb}

The X-ray polarization degree of the northeastern shell of SN~1006 (PD$=22.4\%\pm 3.5\%$)
is smaller than the theoretical maximum value of PD$_{\max}=80\%$ for a  photon index of $\Gamma=2.74$, supporting that the magnetic fields are turbulent. 

The turbulent magnetic fields in the upstream can be generated by streaming instabilities driven by the CR precursors \citep{bell04}. 
The length scale of non-resonant Bell instability for the most rapid growth mode is given as \citep{bell04}:
\begin{multline}
l_{\rm Bell}\sim 2 \times 10^{17}~\cm~\left(\frac{V_{\rm s}}{5\times 10^3~\km\ps}\right)^{-3} \\
\left(\frac{n_0}{0.05~\cm^{-3}}\right)^{-1} \left(\frac{E_{\rm max}}{100~{\rm TeV}}\right)  \left(\frac{B_0}{3~\mu{\rm G}}\right)
\end{multline}
where we take the shock velocity $V_{\rm s}\sim 5\times 10^3~\km\ps$ \citep{katsuda09}, ambient density $n_0 \sim 0.05~\cm^{-3}$ \citep{acero07}, and a maximum energy $E_{\rm max} \sim 100$~TeV \citep{acero10}. 

Assuming the magnetic turbulence in the postshock has similar properties to that in the preshock, we can compare the maximum length scale of the Bell instability with the spatial resolution of IXPE $l_{\rm IXPE}$. 
One expects to observe some highly polarized structures when $l_{\rm Bell} > {\rm min} (l_{\rm X}, l_{\rm IXPE})$, where $l_{\rm X}$ is the synchrotron rim width, and $l_{\rm IXPE}$ is the structure size observed with IXPE.
The FWHM rim width in the 2--7 keV band is $5''$--$48''$ \citep{ressler14}, with an averaged value of $25''$ ($8\times10^{17}~\cm$) for the outer rim and $15''$ ($5\times10^{17}~\cm$) for the inner rim. 
The angular resolution of IXPE is $\sim30''$, corresponding to $10^{18}~\cm$ for SN~1006, but the regions we selected are a few arcminutes in size.
Therefore, the magnetic turbulence scale is smaller than the typical rim width and cannot be resolved with IXPE,  which could explain the relatively low PD.
We caution that the Bell instability is predicted for the preshock region, while magnetic turbulence properties in the postshock are unclear.
Moreover, it is unclear whether the Bell instability can result in radially distributed magnetic fields observed with IXPE.
In a more realistic case, the PD distribution depends not only on the telescope resolution or rim width, but also on the initial turbulence models and the nonlinear physics of turbulent cascades \citep{bykov20}.

The Bell mechanism is not the only process to drive turbulent magnetic fields. Shock interaction with the inhomogeneous medium can lead to magnetic amplification and turbulence in both preshock and postshock regions via turbulent dynamo \citep{giacalone07,beresnyak09,drury12,inoue13,xu17}.
The characteristic scale of the turbulent modification of the magnetic field $l_{\rm RMI}$ depends on the density fluctuation scale $l_{\Delta \rho}$ as \citep{inoue13}: $l_{\rm RMI}\sim l_{\Delta \rho}/(r_c A) $, where $r_c$ is the shock compression ratio, $\Delta \rho$ is the density fluctuation, and $A\simeq (\Delta \rho/\rho)(1+\Delta \rho/\rho)$. 
This scale could be large in SN~1006, compared to $l_{\rm Bell}$, since the remnant evolves in a fairly uniform interstellar medium in the northeast and the south \citep{giuffrida22}.
The northeastern shell of SN~1006 has a very low density of 0.05--0.085~cm$^{-3}$ \citep{acero07,katsuda09}, although there is evidence for denser gas in the northwest with enhanced H$_\alpha$ and FUV emission \citep[e.g.][]{winkler03,korreck04}.
X-ray proper motion measurements found that the northeast shell across different azimuthal angles expands at nearly constant velocities \citep{katsuda09}, which requires a homogeneous ambient medium beyond the shell.

Due to the small density fluctuation, it is likely that the turbulent dynamo is less efficient in SN~1006 than in young SNRs in the clumpy medium. We suggest that CR-induced instabilities could set the magnetic turbulence in SN 1006.

\subsection{Comparison between SNRs observed with IXPE}

\begin{table*}
\caption{X-ray polarization degree, shock velocity and ambient density $n_0$ of Cas~A, Tycho, and SN~1006 northeast
}
\centering
\footnotesize
\begin{tabular}{l|cccccc}
\hline
\hline
& PD (rim) & PD (SNR) & PD (peak) & $V_s$ &  $n_0$ & ref.\\
& (\%) &  (\%) &  (\%) & ($\km\ps$) & (cm$^{-3}$)  &  \\
\hline
Cas~A & $4.5\pm 1.0$ & $2.5\pm 0.5$ & $\sim 15$ & $\sim 5800$ & $0.9\pm 0.3$ & [1][2][3]\\
Tycho & $12\pm 2$ & $9\pm 2$ & $23\pm 4$ & $\sim 4600$ & $\sim 0.1$--0.2 & [4][5][6] \\
SN~1006 NE & $22.4\pm 3.5$ & $\cdots$ & $31\pm 8$ & $\sim 5000$ & $\sim 0.05$--0.085 & [7][8][9][10] \\
\hline
\hline
\end{tabular}
\tablecomments{ The PD values are taken from the nonthermal component. References: [1]-- \cite{vink22},
[2] -- \cite{vink22a}
[3] -- \cite{lee14},
[4] -- \cite{ferrazzoli23},
[5] -- \cite{hughes00},
[6] -- \cite{williams13},
[7] -- this paper,
[8] -- \cite{acero07},
[9] -- \cite{katsuda09},
[10] -- \cite{giuffrida22}
}
\label{tab:compare}
\end{table*}

To date, we have the X-ray polarization measurements for three young SNRs, Cas~A, Tycho, and SN~1006. IXPE has already revealed some interesting differences and similarities among them.

A remarkable similarity is the presence of predominantly radial magnetic fields at the immediate postshock. In contrast, old SNRs show tangential magnetic field orientation \citep{gao11}, which is believed to originate from compressing the tangential components of ambient magnetic fields. 
This reinforces that the magnetic fields in young SNRs probed by X-ray polarimetry are not generated by compressing but need other mechanisms to work.
As discussed in Section~\ref{sec:orientation}, these mechanisms invoke MHD instabilities and efficient CR acceleration that are crucial for understanding CR acceleration mechanisms in SNRs.
But these three SNRs do not necessarily share the origin of their radial magnetic morphology since one mechanism may dominate over others in specific cases.
Compared to Cas~A and Tycho \citep{vink22,ferrazzoli23}, the
radial magnetic morphology of SN~1006 northeast likely favor the efficient 
quasi-parallel CR acceleration in the pre-existing field over the MHD instabilities from density fluctuations.

The three young SNRs show significant differences in the degree of polarization as listed in Table~\ref{tab:compare}.
At the forward shock, the X-ray PD in SN~1006 is around twice of that in Tycho \citep[$12\%\pm 2\%$,][]{ferrazzoli23} and 5 times of that in Cas~A \citep[$4.5\%\pm 1.0\%$;][see Table~\ref{tab:compare}]{vink22}.
This comparison hints that the three young SNRs do not have the same level of turbulent magnetic fields in the immediate postshock region.

\cite{bandiera16} has provided analytic formulae for calculating the polarization parameters of SNRs in turbulent magnetic fields.  Under a simplifying assumption that the magnetic field is a combination of an ordered component $\bar{B}$ and a random component with isotropic Gaussian distribution (with a kernel of $\sigma_B$), the PD is  a function of the turbulence level $\sigma_B/\bar{B}$ and the photon index $\Gamma$ (see their Eq.\ 31). 
For the ``Shell'' region of SN~1006 with a photon index $\Gamma=2.74$,
we obtain the $\sigma_B/\bar{B}=1.2^{+0.2}_{-0.1}$.
We also derived $\sigma_B/\bar{B}\approx 1.8$ in Tycho
rim with $\Gamma=2.82$ and $\sigma_B/\bar{B}\approx 3.3$ for Cas~A in its outer rim with $\Gamma=3.0$.
The comparison shows that the ratio between random and ordered magnetic fields is largest in Cas~A and smallest in SN~1006.
Note that the $\sigma_B/\bar{B}$ values obtained above are based on a simplified assumption of the magnetic-field decomposition.
They have a different meaning in physics from the  magnetic amplification ratio of $10$--$100$ (between the actual field strength in the downstream and ambient field strength) derived from the profiles of the nonthermal X-ray filaments \citep{vink03a}. 

What causes the different turbulent levels among the young SNRs needs more investigation, 
but a possible cause is the different environments.
As discussed in Section~\ref{sec:turb}, SN~1006 is a peculiar case with a tenuous and smooth ambient medium, so the magnetic turbulence is likely purely generated by CR-induced instability.
Cas~A is a young core-collapse SNR evolving in its highly clumpy progenitor winds  \citep{vink96, chevalier03,koo23}.
The diffuse part of the circumstellar medium has an average density of $0.9\pm 0.3$~cm$^{-3}$, while the dense winds contain many knotty structures a few orders of magnitude denser than the diffuse gas \citep{peimbert71,hwang09,lee14,koo23}.
Due to the large density fluctuation, the length scale of turbulence-modified magnetic fields $l_{\rm RMI}$ should be small, and the turbulent dynamo can play an important role in Cas~A.
Tycho has a Type Ia origin, and the average ambient density is 0.1--0.2$~\cm^{-3}$ \citep[e.g.,][]{williams13}.
Nevertheless, the surrounding medium of Tycho is not uniform.
The radio and X-ray proper motion observations of Tycho have revealed that the SNR is experiencing strong deceleration in the northeast and east \citep{reynoso97,williams16,tanaka21}.
The deceleration is suggested to result from a very recent interaction of the SNR with a surrounding molecular bubble swept up by the progenitor winds \citep{lee04,zhou16}.
Therefore, the density fluctuation can cause some magnetic turbulence in Tycho.

Given the aforementioned environmental properties of Cas~A, Tycho, and SN~1006, the MHD turbulent dynamo can work at a decreasing level for these remnants.
This might explain the decreasing magnetic turbulence and thus 
 increasing PD in the three SNRs, although we do not exclude the possibilities that other mechanisms might also play a part and CR-induced instability alone might cause different PDs.
Our IXPE observations suggest that magnetic turbulence, and ultimately CR acceleration in SNRs, is environment-dependent \citep{xu22}.

\subsection{Comparison between the X-ray and radio polarization} \label{sec:radio}

The IXPE observation of \snr\ reveals a possible difference with the radio polarization \citep{reynoso13}.

As shown in Table~\ref{tab:pol}, the average angle of polarization in the radio band is PA$_r=-36\fdg{3} \pm 0\fdg{4}$, while the X-ray value is PA$=-45\fdg{4} \pm 4\fdg{5}$. 
We caution that a uniform RM=$12~\rm{rad~m}^{-2}$ is used to correct the Faraday rotation and the error of RM is not included.
Due to the large uncertainty of RM value \citep{reynoso13} and its influence PA$_r$,
we cannot conclude that there is a large discrepancy in PA between radio and X-ray bands.

X-ray PD ($22.4\%\pm 3.5\%$) in the SNR shell is larger than the radio value ($14.5\%\pm 0.2\%$) at a $\sim 2\sigma$ significance level.
Although the significance is not sufficiently large,
a discrepancy in PD between X-ray and radio measurements is not unexpected.
Firstly, the intrinsic PD of the X-rays can be larger than that of the radio emission. 
The maximum polarization PD$_{\rm max}=\Gamma/(\Gamma+2/3)$ is 80\% in the X-ray band with $\Gamma=2.74$ and 70.5\% in the radio band with $\Gamma=1.6$ \citep{reynoso13}.
Secondly, the X-ray emission is radiated from a smaller layer than the radio emission, and thus may have less depolarization.
The synchrotron emissivity profile reflects the strength of the magnetic fields and particle distribution.
The X-ray synchrotron emission originates from electrons freshly accelerated to TeV energy, which lose their energy $E$ quickly with a characteristic timescale of $\tau_{\rm loss} = 640/(B^2E)$~s, with the electron energy, E, in units of erg and the magnetic field strength, B, in units of Gauss \citep[e.g.,][]{reynolds98}.
Radiative energy losses as particles advect downstream can shrink the rim width in the higher energy band
$l_{\rm loss}\approx v_d \tau_{\rm loss}$, with $v_d$ the fluid velocity downstream of the shock.
The rim width might also be regulated by the diffusion transport of particles or/and magnetic damping, but \cite{ressler14} compared with the observations and supported that the filament widths in SN~1006 decrease with increasing photon frequency $\nu$ (width$\propto \nu^{m_E}$, $m_E\sim -0.3$ to $-0.8$).
With a larger emitting volume than the X-rays, the radio emission also tends to be depolarized as the polarization angles vary along the line of sight.
Thirdly, if the RM is nonuniform but we incorrectly used a uniform RM, the calculated PA$_r$ value may vary across the shell. Consequently, combining the radio polarization data from a large region might cause a depolarization.

\section{Summary}

In this article, we report the first X-ray polarization imaging observations of the northeastern shell of SN~1006. The main results are summarized as follows:

\begin{enumerate}
    \item We show the spatial distribution of PD and PA of the shell.
   The PA in different pixels shows a tangential distribution along the shell, suggesting that the overall magnetic orientation parallels the shock normal or has a radial distribution.

    \item The overall degree of polarization in the northeastern shell is $22.4\%\pm 3.5\%$ and PA$=-45.5^\circ\pm 4.5^\circ$ (background removed).
    We checked the variation of polarization properties by separating the shell into a few regions, but did not found significant differences of PD and PA across the shell.

    \item The magnetic field orientation in SN~1006 supports the efficient quasi-parallel CR acceleration. 

    \item We compared X-ray polarization measurements of Cas~A, Tycho, and SN~1006. The remarkable similarity is the radially distributed magnetic fields. But they also show an increasing PD, likely reflecting a decreased turbulent level of the magnetic fields. We compared their environmental density and interpreted that the different density and density fluctuations could cause the PD discrepancy between the three SNRs. The IXPE observations thus support that magnetic turbulence is environment-dependent.

    \item The X-ray PD in the shell is larger than the radio PD ($14.5\%\pm 0.2\%$) at a $\sim 2\sigma$ level.
    The possible explanation includes a larger intrinsic PD in X-rays, a larger radio emission emitting area that causes a depolarization, or a depolarization due to applying incorrect RM distribution.

\end{enumerate}

\begin{acknowledgements}
The Imaging X-ray Polarimetry Explorer (IXPE) is a joint US and Italian mission. The US contribution is supported by the National Aeronautics and Space Administration (NASA) and led and managed by its Marshall Space Flight Center (MSFC), with industry partner Ball Aerospace (contract NNM15AA18C). The Italian contribution is supported by the Italian Space Agency (Agenzia Spaziale Italiana, ASI) through contract ASI-OHBI-2017-12-I.0, agreements ASI-INAF-2017-12-H0 and ASI-INFN-2017.13-H0, and its Space Science Data Center (SSDC), and by the Istituto Nazionale di Astrofisica (INAF) and the Istituto Nazionale di Fisica Nucleare (INFN) in Italy. This research used data products provided by the IXPE Team (MSFC, SSDC, INAF, and INFN) and distributed with additional software tools by the High-Energy Astrophysics Science Archive Research Center (HEASARC), at NASA Goddard Space Flight Center (GSFC).
P.Z.\ acknowledges the support from NSFC grant No.\ 11590781 and Nederlandse
Organisatie voor Wetenschappelijk Onderzoek (NWO) Veni Fellowship, grant no.\ 639.041.647.
C.-Y. Ng and Y. J. Yang are supported by a GRF grant of the Hong Kong
Government under HKU 17305419.
N.B. was supported by  the INAF MiniGrant  ``PWNnumpol - Numerical Studies of Pulsar Wind Nebulae in The Light of IXPE''.
\end{acknowledgements}

\software{
ixpeobssim (vers.\ 30.0) \citep{baldnini22},
DS9 \citep{joye03,sao00},
SAS \citep{sas14},
Xspec \citep[vers.\ 12.9.0u,][]{arnaud96}.
}

\begin{appendix} 
\label{sec:region2}

Figure~\ref{fig:spec} shows the IXPE DU1 spectrum of the region ``Shell'', with the background spectrum subtracted. The solid line shows the modeled background spectrum.
Above $\sim 4$~keV, the background flux becomes comparable to or brighter than the source flux.
Therefore, we select only the 2--4 keV energy band to examine the polarization properties. 

\begin{figure}
\centering
	\includegraphics[width=0.5\textwidth, angle=0]{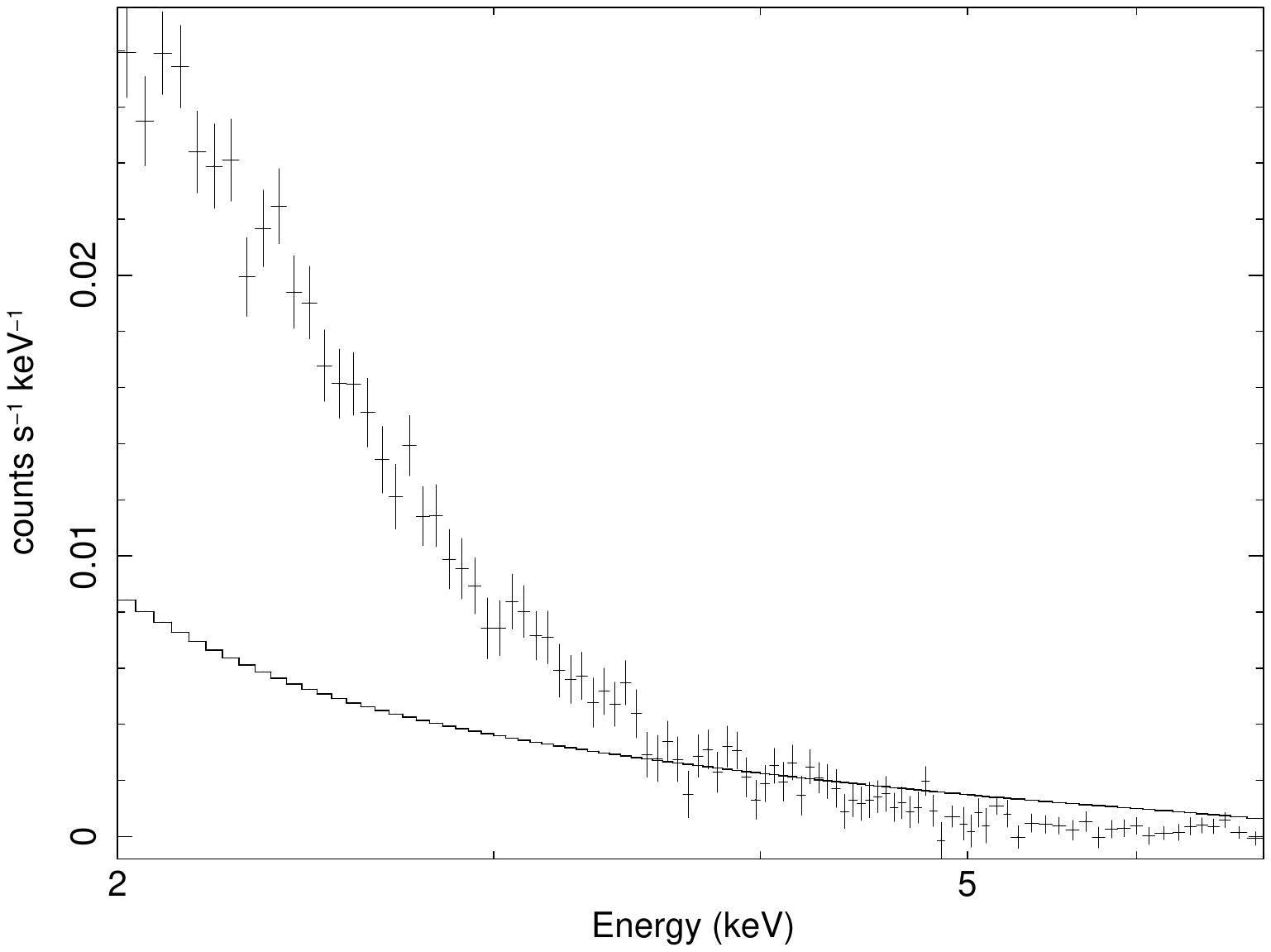}
    \caption{
   Background-subtracted Stokes I spectrum from DU1 (data points) of the region ``Shell'' and the background level (solid line). }
    \label{fig:spec}
\end{figure}

\begin{figure}
\centering
	\includegraphics[width=0.5\textwidth, angle=0]{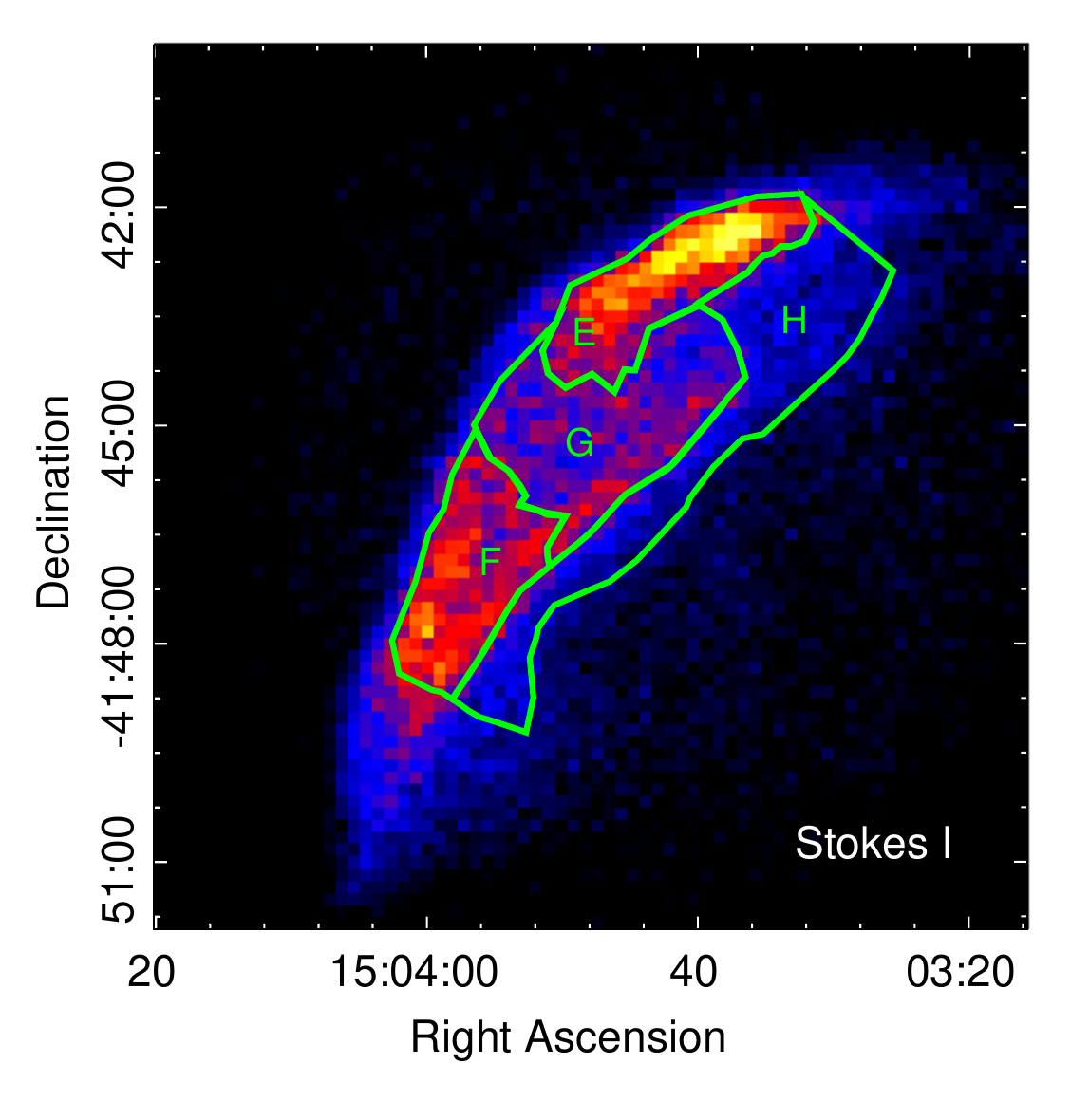}
    \caption{
   Stokes I image overlaid with regions E--H with decreasing brightness. }
    \label{fig:region2}
\end{figure}

\begin{table}
\caption{X-ray polarization results of 4 regions with different X-ray brightness
}
\centering
\begin{tabular}{l|ccc}
\hline
\hline
Region & PD  &  PA & $\sigma$ \\
 & (\%) & ($^\circ$) & \\
 \hline
E &  $20.5\pm 5.0$ & $-47.1 \pm 6.9$  & 4.1 \\
F & $22.5\pm 5.4$ & $-36.2 \pm 6.9 $ & 4.2 \\
G & $21.7\pm 5.6$ & $-45.6\pm 7.4$ & 3.9 \\
H & $26.0\pm 7.7$ & $-39.0\pm 8.4$ & 3.4 \\
\hline
\end{tabular}
\tablecomments{The regions are denoted in Figure~\ref{fig:region2}. 
The polarization errors are provided at the 1$\sigma$ confidence level.}
\label{tab:pol2}
\end{table}

To investigate whether the polarization parameters vary with the X-ray surface brightness, we calculated the background-subtracted PD and PA in four new regions, E, F, G, and H, with decreasing X-ray brightness (see Figure~\ref{fig:region2}). 
The obtained results are shown in Table~\ref{tab:pol2}.
The PD and PA values of the four regions are consistent with each other, suggesting that PD is near uniform across the shell and independent of the X-ray surface brightness.

\end{appendix}

\end{document}